\documentclass[a4paper,fleqn]{article}

\usepackage{graphicx}

\usepackage[font = scriptsize, format=hang]{caption}

\usepackage{multicol}        
\title{Direct numerical simulation of a compressible multiphase
  flow through the fast Eulerian approach}

\author{Matteo Cerminara
  \\
  Scuola Normale Superiore di Pisa
  \\
  Istituto Nazionale di Vulcanologia e Geofisica
  \\
  Pisa, ITALY, matteo.cerminara@sns.it 
  \and
  Luigi C. Berselli
  \\
  Dipartimento di Matematica
  \\
  Universit\`a di Pisa
  \\
  Pisa, ITALY, berselli@dma.unipi.it 
  \and
  Tomaso Esposti Ongaro
  \\
  Istituto Nazionale di Geofisica e Vulcanologia
  \\
  Section of Pisa
  \\
  Pisa, ITALY, ongaro@pi.ingv.it
  \and
  Maria Vittoria Salvetti
  \\
  Dipartimento di Ingegneria Aerospaziale
  \\
  Universit\`{a} di Pisa
  \\
  Pisa, ITALY, mv.salvetti@ing.unipi.it
  }

\begin{document}

\maketitle

\section{Introduction}
Our work is motivated by the analysis of ash plume dynamics, arising in the study of volcanic eruptions. 
Such phenomena are characterized by large Reynolds number (exceeding $10^7$) and a large number of polydispersed particles~[1].
Thus, the choice of the methodology to be used is straightforward: we need LES of a multiphase gas-particles flow.
Since the simulation of the behavior of a large number of dispersed particles is very difficult with Lagrangian
methods, we model the particles as a continuum, Eulerian fluid (dust), by using
reduced models involving two fluids, as proposed in Ref.~[2,3,4]. Moreover, we need a robust numerical scheme to
simultaneously treat compressibility, buoyancy effects and turbulent dispersal dynamics.

\begin{figure}[htb]
\centerline{
\includegraphics[width=0.95\columnwidth]{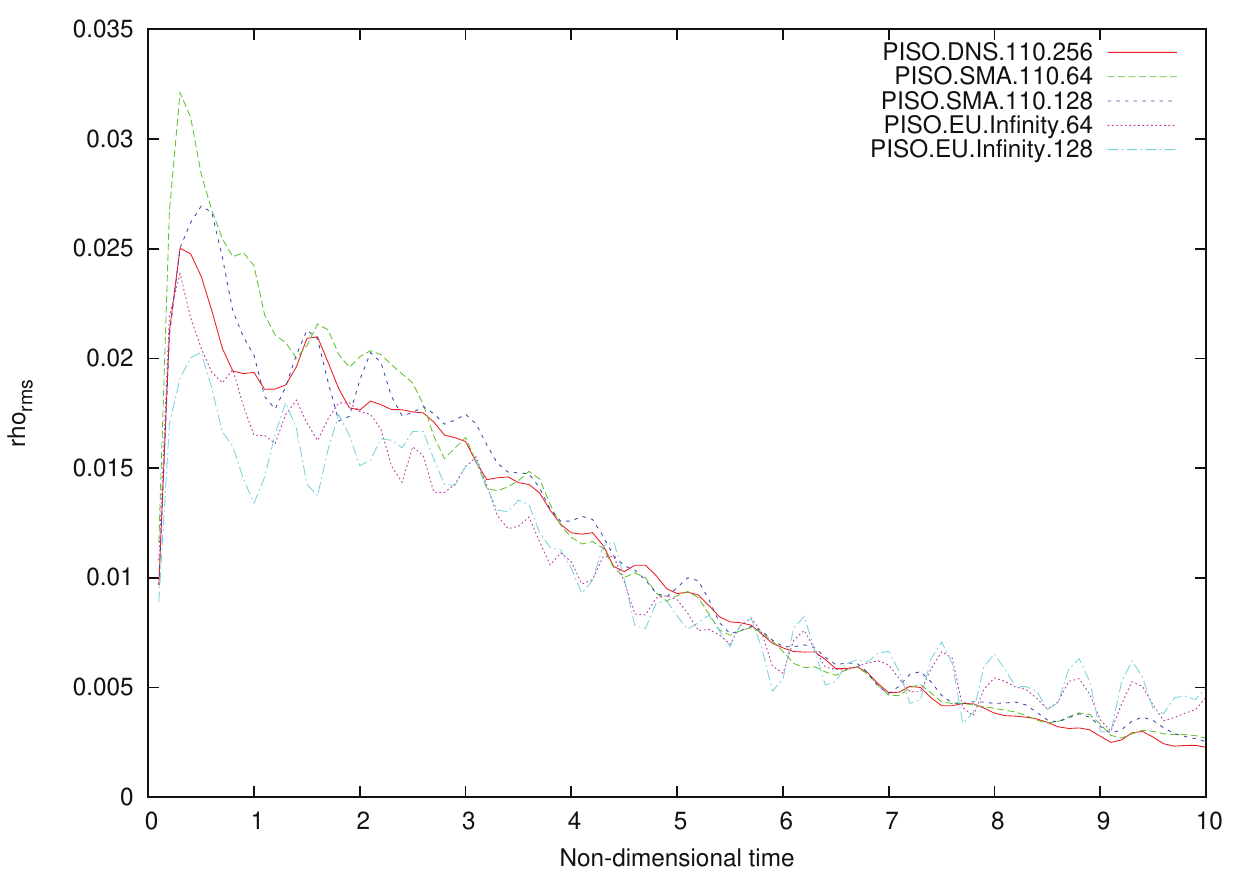}}
\caption{Non dimensional density r.m.s., where the r.m.s. Mach number is $0.2$  and $t=\pi$ is one initial large-eddy turnover time.\textsuperscript{1}}
\label{fig:rho}
\end{figure}

We analyze the turbulence properties of such models in a homogeneous and isotropic setting, with the aim of formulating a LES model.  In particular, we examine the development of freely decaying homogeneous and
isotropic turbulence in subsonic regime
(the r.m.s. Mach number either 0.2 or 0.5, see Fig.~\ref{fig:rho}) using
OpenFOAM\textsuperscript{\textregistered}, which is one of the best
known CFD open source software packages.


\section{Results}

\begin{figure}[htb]
\centering
\includegraphics[width=0.45\columnwidth]{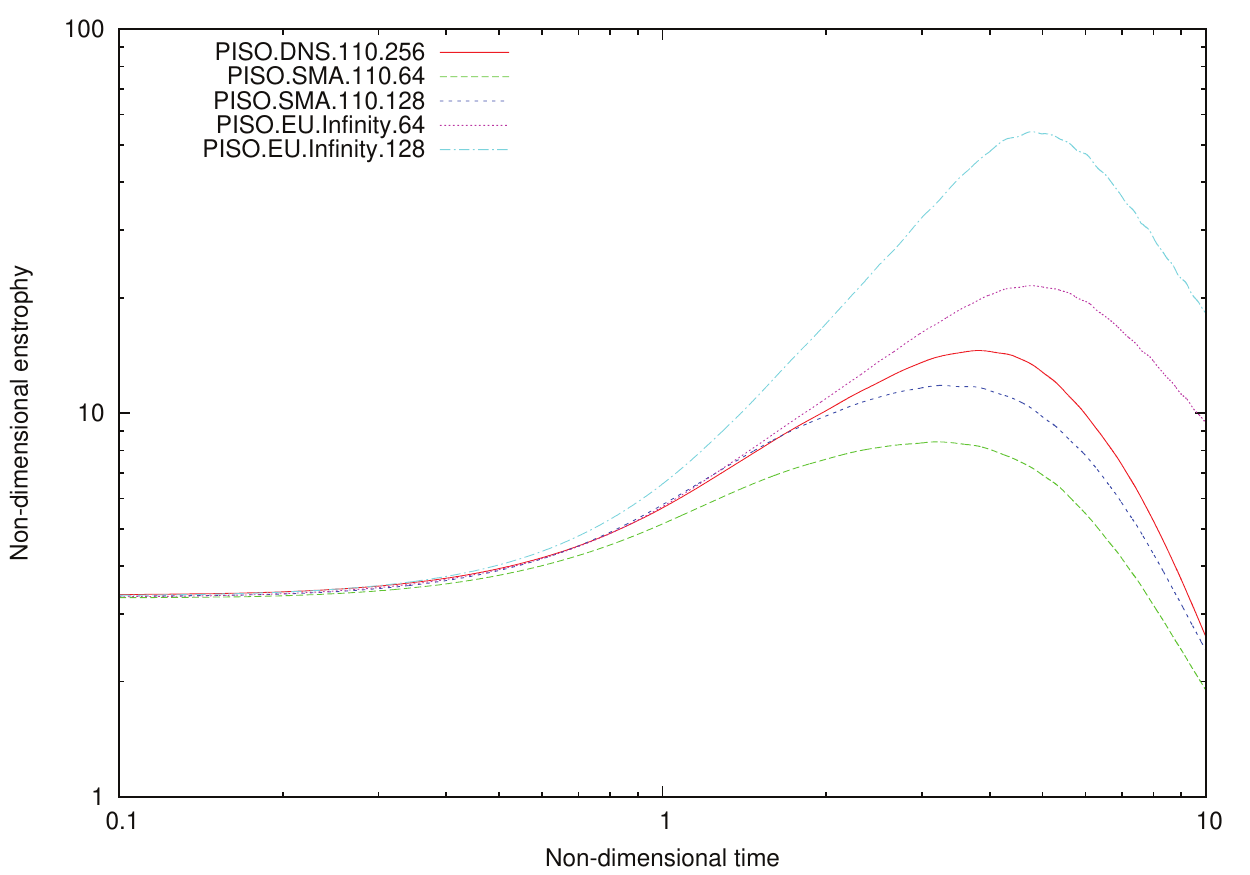}\,
\includegraphics[width=0.45\columnwidth]{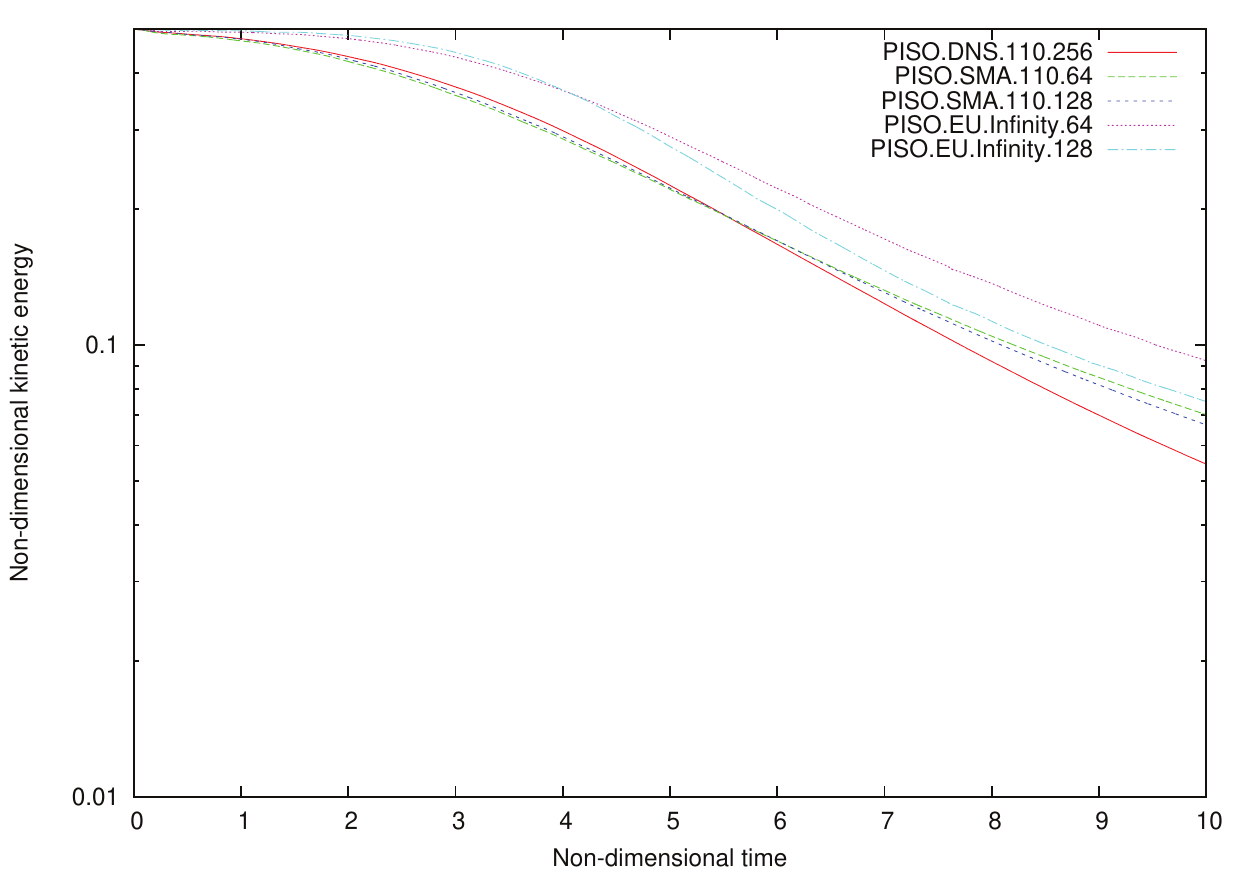}
\caption{Enstrophy and kinetic energy evolution. Cases are the same of Fig.~\ref{fig:rho}.}
\label{fig:enstrophy}
\end{figure}

\begin{figure}[htb]
\centerline{
\includegraphics[width=0.95\columnwidth]{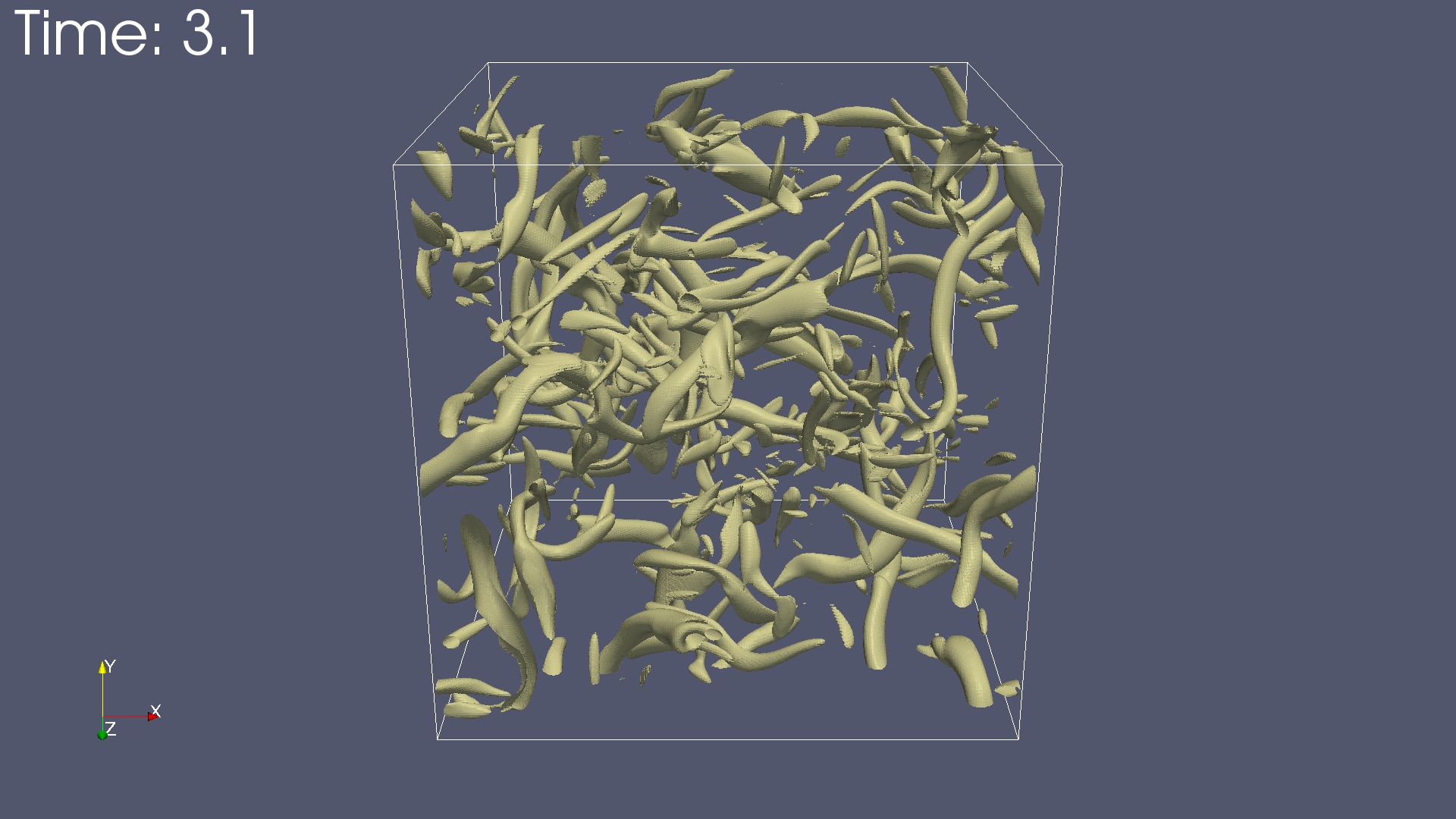}}
\caption{Homogeneous and isotropic turbulence as obtained with OpenFOAM\textsuperscript{\textregistered} (simulation PISO.DNS.110.256\textsuperscript{1}, $M_\textup{rms}=0.2$). Here iso-surfaces of the second invariant $Q$ are represented.}
\label{fig:HmIs}
\end{figure}

The first goal is to appraise the capabilities of the adopted pressure based scheme
for the mono-phase fluid. This is obtained by comparing our DNS and LES
with results available in the literature (cf. e.g. Ref.~[5]). In particular, we
use the PISO solver (see e.g.~[6]) with a fourth order central interpolation scheme and an adaptive time-step~[7].
The code has a behavior comparable to that discussed in~[5] and in particular it reproduces well the energy spectrum, the enstrophy and
kinetic energy (Fig.~\ref{fig:enstrophy}), and the Taylor micro scale evolution.
The typical elongated structures of homogeneous turbulence are
also reproduced, see Fig.~\ref{fig:HmIs}.

We then implemented in the OpenFOAM's \texttt{C++} library a new solver for an Eulerian two-phases gas-particles model and also a
fast Equilibrium-Eulerian model~[3], both in the Fourier\footnote{In the energy balance equation, the heat flux is proportional to the temperature gradient: $$ q = - k \nabla T\,. $$} and the barotropic case\footnote{If the thermodynamic transformation is approximatively isotropic, it is unnecessary to solve the energy conservation low. It is sufficient to take the pressure proportional to the density to the power of the specific heat ratio: $$ p \propto \rho^\gamma\,. $$}. 
In particular, the Equilibrium-Eulerian model is a first-order approximation of the particles
momentum balance equations, using the Stokes law and a perturbative method valid when the Stokes number approaches zero. In other words, we have cinematic quasi-equilibrium between the gas and the particle phase.

The latter model has the advantage to be as fast as the Marble's Dusty Gas model\footnote{Zero-order version of the Equilibrium-Eulerian model, cf.~[2].}, and it seems at the same time capable to capture important phenomena like preferential concentration~[3,4].

We then performed simulations of freely decaying two-phase homogeneous
and isotropic turbulence, comparing the results with those of DNS in which the particle dynamics is computed through a Lagrangian approach, e.g.~[8]. The aim of the present work is twofold. First, it will give an appraisal of the capabilities of the fast Eulerian approach to accurately capture the particle dynamics, in particular, for particle inertia and Mach number values of interest in the target geophysical application. Second, the DNS data base will be used in a-priori tests to get indication on the sub-grid scale (SGS) terms which must be modeled in LES combined with the fast Eulerian approach, and, eventually, to develop ad hoc model to be used in LES simulations of more realistic and complex flow configurations.

The results of the DNS simulations, which are still in progress,  will be shown and analyzed in the final contribution.

\section{References}

\noindent
[1] Valentine, G. A. (2001). Eruption column physics. {\em From Magma to Tephra: Modelling Physical Processes of Explosive Volcanic Eruptions} (edited by A. Freundt and M. Rosi). Elsevier, New York, 91--136.

\noindent
[2] Marble, F.E., (1970) Dynamics of Dusty Gases. {\it Ann. Rev. Fluid
  Mech.} {\bf 2}, 397--446.

  \noindent
[3] Ferry, J. \& Balachandar, S.,   (2001). A fast Eulerian method for disperse
two-phase flow   {\em Int. J. Multiphase Flow} {\bf 27}, 1199-1226.

\noindent
[4] Balachandar, S. \& Eaton, J. K. (2010). Turbulent Dispersed
Multiphase Flow.  {\em Ann. Rev. Fluid Mech. } {\bf 42}, 111--133.

  \noindent
[5] Garnier, E., Mossi M., Sagaut, P., Compte, P. \& Deville,
M. (1999). On the use of Shock capturing schemes for Large-Eddy
Simluation. {\it J. Comp. Phys.} {\bf 153} 273--311.

\noindent
[6] Demirdzic, I., Lilek, Z. \& Peric, M. ()1993). A collocated finite volume method for predicting flows at all speeds. {\em Int. J. Numer. Methods Fluids}, {\bf 16}, 1029--1050.
 
 \noindent
[7] Kay, D. A., Gresho, P. M., Griffiths, D. \& Silvester,
D. J. (2010). Adaptive time-stepping for incompressible flow. Part II:
Navier-Stokes equations.  {\it SIAM J. Sci. Comput.} {\bf 32}, 111--128.

\noindent
[8] Biferale, L., Boffetta, G., Celani, A., Devenish, B. J., Lanotte, A., \& Toschi, F. (2005). Lagrangian statistics of particle pairs in homogeneous isotropic turbulence. {\em Physics of Fluids}, {\bf 17}(11), 1-9.



%

\footnotetext[1]{Simulations names are organized as follows: ``scheme''.``sub-grid model''.``maximum wave number''.``number of cells$^{1/3}$''; thus {\bf PISO.noM.110.128} stands for a simulation done with a PISO scheme, without sub-grid model, where the maximum wave number is 110 ($Re_{k_0} \simeq 210$) in a box with $128^3$ cells.}


\newpage

\end{document}